# The Amsterdam Manifesto on OCL

Steve Cook[1], Anneke Kleppe[2], Richard Mitchell[3], Bernhard Rumpe[4], Jos Warmer[5], and Alan Wills[6]

[1] IBM European Object Technology Practice, UK,
http://www.ibm.com/
sj_cook@uk.ibm.com

[2] Klasse Objecten, NL-3762CT Soest, Netherlands,
http://www.klasse.nl/
A.Kleppe@klasse.nl

[3] University of Brighton, Brighton BN2 4GJ, UK,
http://www.it.brighton.ac.uk/staff/rjm4/
Richard.Mitchell@brighton.ac.uk

[4] Technische Universität München, Institut für Informatik, 80290 München,
http://www.in.tum.de/~rumpe/
Bernhard.Rumpe@in.tum.de

[5] Klasse Objecten, NL-3762CT Soest, Netherlands,
http://www.klasse.nl/
J.Warmer@klasse.nl

[6] TriReme International, Manchester, UK,
http://www.trireme.com/
alan@trireme.com

**Abstract.** In November 1998 the authors participated in a two-day workshop on the Object Constraint Language (OCL) in Amsterdam. The focus was to clarify issues about the semantics and the use of OCL, and to discuss useful and necessary extensions of OCL. Various topics have been raised and clarified. This manifesto contains the results of that workshop and the following work on these topics. Overview of OCL.

## 1  Overview of OCL

The Object Constraint Language (OCL) is a textual specification language, designed especially for the use in the context of diagrammatic specification languages such as the UML [Booch98]. The UML offers several kinds of diagrams, dedicated to describe different aspects of a system, such as structure, interaction, state based behaviour or deployment. However, when regarding the UML diagrams as a language, it turns out that the diagram-based UML is limited in its expressiveness. Although the UML is powerful and covers many important situations, it is often not sufficient to describe certain important constraints. Using natural language on the one hand introduces freedom of misinterpretations and on the other hand gives tools no chance to cope with it.

Therefore the Object Constraint Language was introduced as a textual add-on to the UML diagrams. OCL is deeply connected to UML diagrams, as it is used as textual addendum within the diagrams, e.g. to define pre- and postconditions, invariants, or transition guards, but also uses the elements defined in the UML diagrams, such as classes, methods and attributes. The following example taken from the Journal of Object-Oriented Programming [Warmer99b] shows some important features of the OCL and how to apply them in class diagrams.

### 1.1   Example of the Use of Constraints

In this article we will use the example class diagram from Figure 1. It shows a model of a hotel. The hotel has a number of rooms that can be rented by guests. There are also a number of bathrooms, which are either connected to a specific room, or are used to service multiple rooms on the floor. The modelled hotel has a number of business rules that are very common to hotels:

- The number of guests in each room doesn't exceed the number of beds in the room.

- The hotel allows one extra bed for a child of up to 4 years old per room.

- All guests in each room need are officially registered in the hotel.

- The bathroom connected to a room may only be used by guests of that room.

- Bathrooms not connected to a room may only be used by guests of rooms on the same floor.

Examining the example class diagram, we find that many of the common business rules for hotels can not or at least not easily be expressed in the diagram. We will show in the following how to express these business rules as OCL constraints on the class diagram. This will keep the diagram simple and easy to understand while adding details to the model under construction.

Using the size property on a collection also allow us to describe the first business rule as an invariant constraint on the association from Room to Guest. There can be no more guests in a room than there are beds (business rule number 1).

```
context Room invariant:
guests->size <= numberOfBeds
```

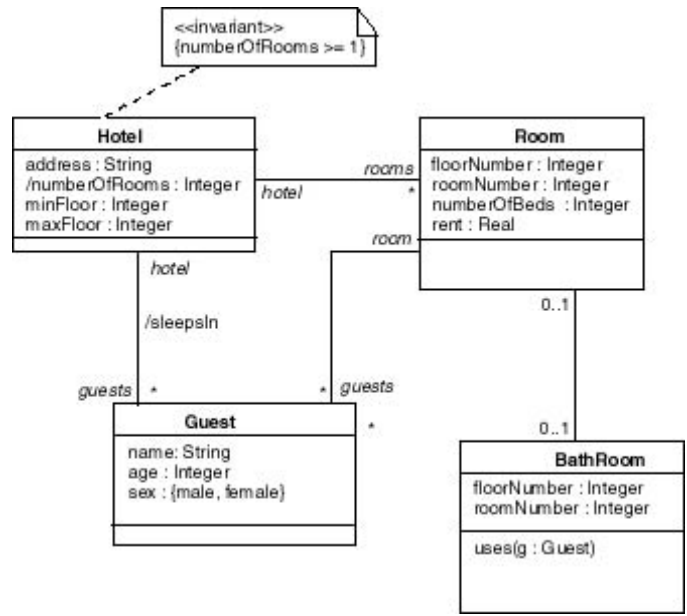

**Fig. 1.** Example class diagram

This kind of constraint on the multiplicity of an association occurs quite often in practice. The UML class model specifies multiplicity 'zero to many' in cases where the actual upper (or lower) bound of the multiplicity is not known beforehand, but depends on the value of an attribute of an object at runtime.

The *exists* operation evaluates to true if the argument expression evaluates to true for at least one element in the collection. Using this we can rewrite the constraint stated above about the number of guests allowed in a room. The hotel in the example allows for an extra bed for a child of four years at most (business rule number 2). Therefore the constraint given above must be adjusted to:

```
context Room invariant:
guests->size <= numberOfBeds or
( guests->size = numberOfBeds + 1 and
  guests->exists(g : Guest | g.age <= 4)
)
```

The *collect* operation results in a new collection. The elements of the resulting collection are the values of the argument expression for each element in the original collection. An example of an invariant using *collect* is shown for Hotel. This constraint expresses our business rule number 3:

```
context Hotel invariant:
guests = rooms->collect(guests)->asSet
```

For the hotel in the example the business rules have been established that a guest can only use a bathroom when it is either attached to his/her room, or when it is not attached to a room and located on the same floor as his/her room (number 4 and 5). These business rules are specified in the precondition of the operation *uses* of Bathroom. The number of uses is counted in attribute *usage* (that we deliberately omitted in the class diagram):

```
context Bathroom::uses(g : Guest)
pre: if room->notEmpty then
         room.guests->includes(g)
     else
         g.room.floorNumber = self.floorNumber
     endif
post: usage = usage@pre + 1
```

### 1.2   Status of OCL

OCL has syntactic similarities to Smalltalk (amongst others [Lalonde90]). OCL could also be given a Java/C++-like syntactic flavour without affecting the usefulness of OCL. Syntactic flavour is to a large extent a matter of taste. However, more important are the concepts within the language. Here we name the most important features of OCL. Please, be reminded that this manifesto is not an introduction to OCL, this can be found e.g. in [Warmer99a]:

- Tight integration with the UML notation
- Pure specification language without operational elements
- Basic types like Boolean, Integer, Float
- Container types Collection, Bag, Sequence, and Sets with appropriate operators
- The full typesystem also includes the types induced by class definitions within UML
- Navigation expressions to navigate along associations with various multiplicities
- Boolean operators to allow full propositional logic
- Existential and universal quantifiers over existing sets of objects
- Specifications are (to a large extent) executable and can therefore be used as assertions when animating the UML diagrams or generating code

OCL is a specification language that tries to mediate between the practical users needs and the theoretical work that has been done in that area. In particular much of the theoretical work was done in the areas of algebraic specification languages, such as Spectrum [Broy93-1, Broy93-2], CIP [Bauer85], ACT ONE [Ehrig85], ACT TWO [Ehrig90], TROLL [Conrad92, Hartmann94, Jungclaus91], logic systems such as HOL [Gordon93, Paulson94], LCF [Gordon79, Regensburger94]. Other theoretical work worth mentioning is on model based specification techniques such as Z [Spivey88, Spivey89] and VDM [Jones90], functional programming languages such as Gofer [Jones93], Haskell [Hudak92] and ML [Paulson91], object-oriented methods such as Syntropy [Cook94], and also the data base query languages such as SQL [Date87].

The best ideas from these areas, such as navigation expressions or container types as abstract data types, have been taken and combined into a language that is dedicated to software engineers. OCL does not only have a syntax similar to Smalltalk, but also provides expressive operator names to increase readability of OCL constraints.

The current status of OCL is as follows:

The official OCL specification in version 1.3 has been published by the OMG together with the UML specification in version 1.3.

Parsers for OCL specifications, written in Java are available at:

http://www.software.ibm.com/ad/ocl
http://www.db.informatik.uni-bremen.de/umlbib/
http://www-st.inf.tu-dresden.de/UMLToolset/

A comprehensive list of publications can also be found at http://www.uni-bremen.de/umlbib/.

In [Rumpe98] we have discussed that there exist several degrees of formality of a notation. If the syntactic shape of a notation is precisely defined, e.g. for OCL a grammar is given, then the syntax is formalised. However, based on the syntax the meaning of the notation has still to be defined.

OCL does currently not have a formally defined meaning and can therefore only be regarded as semi-formal. Due to the tight connection of OCL with the UML diagrams, the definition of a formal semantics for OCL must be based on a formal semantics for the UML. This is a difficult task.

### 1.3 Contents of This Paper

The manifesto documents the results of a two-day workshop held in Amsterdam discussing various topics and issues of the OCL. The results have been classified roughly in four groups: bug fixes; clarifications; extensions; applications. Some of the issues discussed below do not fall clearly into one of these categories, but belong to several groups. E.g. it is sometimes necessary to propose an extension in order to fix a bug in the language. The following four chapters are structured along this four groups. Each discussed topic corresponds to one section.

## 2   Bug Fixes of the Language

### 2.1     Remove Direct Type

In UML 1.1 the standard operation "oclType" would result in the type of an object. Because objects can have multiple types, this operation was ill defined. In UML 1.3 this operation has been removed. The operations "oclIsTypeOf" and "oclIsKindOf" can be used instead.

### 2.2     Adaptation of OclAny

OCL provides a special type, called "OclAny". This type is meant to be a common supertype of all types. The specification of OclAny in the UML 1.1 and 1.2 definitions was inconsistent. It includes the basic types, such as Boolean, types defined in UML diagrams, like Flight, and collection types, like Set(Flight). In particular Set(OclAny) is again a type included in OclAny. Although, there are type systems dealing with such a situation, these type systems are rather complex. Also for practical purposes this complication is not necessary.

To remedy this situation, the type OclAny was adapted in UML 1.3 to include only the basic types of OCL and the types defined within the UML diagrams. In particular, none of the collection types, e.g. Set(Flight) or Sequence(Boolean), are subtypes of OclAny. It is still possible to build and work with Set(OclAny), but this type is not included in OclAny. Please note, that there is also no inclusion in the other direction, although elements of OclAny could be regarded as one element sets, bags, or sequences as in OBJ [Goguen92].

The adaptation of OclAny leads to an improved and simplified type system. Using this fix of the type system and ignoring the two meta-types OclExpression and OclType, we get a strong type system in the sense, that type checking can be done within the parser.

### 2.3     "allInstances" Considered Dangerous

The definition of "allInstances" is problematic. It is unclear in what context all instances should be taken. It is much better style and much clearer to make sure that any collection of objects is only reached by navigation.

The following paragraph has been added to the UML 1.3 OCL specification to warn users of the potential pitfalls of using allInstances. The future use of OCL will show, whether this is sufficient or a more relased use e.g. of Integer.allInstances will be needed.

From theory we know that only the use of the existential and universal quantifier over infinte sets, like Integers, make a logic language first order, otherwise it is a propositional logic only.

**NB:** The use of `allInstances` has some problems and its use is discouraged in most cases. The first problem is best explained by looking at the types like Integer, Real and String. For these types the meaning of `allInstances` is undefined. What does it mean for an Integer to exist? The evaluation of the expression `Integer.allInstances` results in an infinite set and is therefore undefined within OCL. The second problem with `allInstances` is that the existence of objects must be considered within some overall context, like a system or a model. This overall context must be defined, which is not done within OCL. A recommended style is to model the overall contextual system explicitly as an object within the system and navigate from that object to its associated instances without using `allInstances`.

## 3    Clarifications of the Syntax and Semantics of OCL

### 3.1    Introduction to the Boolean Operators and Undefined Expressions in OCL

OCL provides the following operators with Boolean arguments and/or results:
  =, not, and, or, xor, implies, if-expression.
This section presents informal definitions of the operators, aimed at users of OCL. The definitions are intended for incorporation into the next release of the definition of OCL. It in fact turns out that also the given definitions are intended to give an intuitive explanation, the characterisation of an operation through truth-tables and through reduction to known operations, as used below, is a fully precise definition.
The definitions are presented in two parts. First, the meanings of the operators are given using truth tables and equations. Then there is a short discussion of the use of Boolean operators within OCL constraints.
In UML, and hence in OCL, the type Boolean is an enumeration whose values are "false" and "true", and a Boolean expression is one that evaluates to a Boolean value.
In what follows, b, b1 and b2 are expressions of type Boolean.

#### 3.1.1    The = Operator

Two Boolean expressions are equal if they have the same value. The following table defines the equality operator. Thus equality can be used, for instance, to denote the equivalence of two properties.

| b1    | b2    | b1 = b2 |
|-------|-------|---------|
| false | false | true    |
| false | true  | false   |
| true  | false | false   |
| true  | true  | true    |

### 3.1.2    The NOT Operator

The not operator is defined by the following table.

| b | not b |
|---|---|
| true | false |
| false | true |

### 3.1.3    The AND Operator

The and operator is commutative, so that

    ( b1 and b2 ) = ( b2 and b1 )

It is defined by the following two equations.

    ( false and b ) = false
    ( true and b ) = b

Applying the above commutativity rule to the equations, we get:

    ( b and false ) = false
    ( b and true ) = b

and therefore the following table holds:

| b1 | b2 | b1 and b2 |
|---|---|---|
| false | false | false |
| false | true | false |
| true | false | false |
| true | true | true |

### 3.1.4.    The OR Operator

The or operator is also commutative, so that

    ( b1 or b2 ) = ( b2 or b1 )

It is defined by the following two equations.

    ( false or b ) = b
    ( true or b ) = true

Once again, it is possible to apply the commutativity rule to the defining equations to produce two more equations:

    ( b or false ) = b
    ( b or true ) = true

and a truth table:

| b1    | b2    | b1 or b2 |
|-------|-------|----------|
| false | false | false    |
| false | true  | true     |
| true  | false | true     |
| true  | true  | true     |

### 3.1.5   The XOR Operator

The xor operator ("exclusive or") holds if exactly one of its arguments holds. It is therefore similar to the or operator, but excludes the case that both arguments are true. The xor operator is commutative, so that
    ( b1 xor b2 ) = ( b2 xor b1 )
It is defined in terms of the and, or and not operators.
    ( b1 xor b2 ) = ( (b1 or b2) and not (b1 and b2) )

### 3.1.6   The IMPLIES Operator

The implies operator allows us to formalise statements of the following kind:
"if b1 is true then b2 must also be true (but if b1 is false we don't say anything about b2)".
Such a statement can be formalised by the OCL expression

    b1 implies b2

which constrains b2 to be true whenever b1 is true. The implies operator is defined by the following equations:

    ( false implies b ) = true
    ( true implies b ) = b

It follows that the expression "b1 implies b2" can be false exactly if b1 is true and b2 is false.

### 3.1.7   The IF-Expression

An if-expression takes the form

    if b then e1 else e2 endif

in which b is a Boolean expression and e1 and e2 are OCL expressions of compatible types. If b is true, the value of the whole expression is e1. If b is false, the value of the whole expression is e2.

In contrast to an if statement in procedural languages, an if expression has a value. Therefore the else clause is always necessary. The resulting type of the if-expression is the least type T such that the types of both argument expressions e1 and e2 conform to that type. (See also the type conformance rules of OCL). However, if multiple inheritance is allowed, there need not be one least type T (but more such types). In this case a typechecker needs help through explicit type information for the arguments or the result.

Here is a small example. Assume that count is an integer variable and x is a real variable. The literals 100 and 0 are Integers. The if-expression

if (count <= 100) then x/2 else 0 endif

has type real (because e1 and e2 are of type real and integer, respectively, and integer conforms to real). The value of the whole expression is (x/2) if (count<=100). The value of the whole expression is the real number 0 if (count>100).

### 3.1.8   Boolean Operators and Undefined Expressions

The meanings of the Boolean operators have been presented in the preceding subsections. The meanings have been chosen to support the intuitions of those who write and read OCL constraints. Sometimes it is necessary to write expressions of the form

"if b1 is true, b2 should also be true"

even though we know that "b2" has no meaning when "b1" is false.

Here is an example, based on a fragment of a model of a library system, which has two associations between class Title and class Reservation.

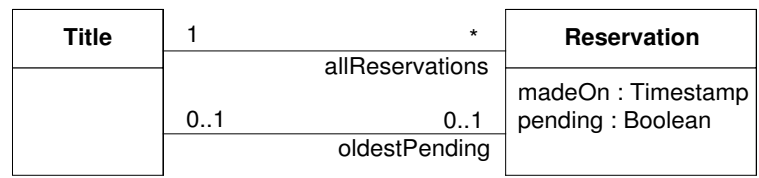

To reserve a title means to put in a request to borrow any copy with that title. A reservation object is created as part of the act of reserving a title. If one or more of the reservations for a particular title is pending then there must be an "oldestPending", which is the reservation that is next in line to be satisfied when a copy with the right title becomes available. (If there are no pending reservations, there is no "oldestPending" reservation, and "allReservations" captures only historical information.)

There is also a Timestamp class, and we are interested in one of its methods.

| Timestamp |
|---|
| notAfter( other : Timestamp ) : Boolean |

The "notAfter" query returns true if the receiver is a timestamp that is at the same time as, or earlier than, the "other" timestamp.

Here is an invariant to define the intended meaning of the "oldestPending" association. The OCL definition includes a comment to explain the formal part.

```
context t: Title invariant:
  t.allReservations->select(pending)->isEmpty implies
      t.oldestPending->isEmpty
and
  t.allReservations->select(pending)->notEmpty implies
      ( t.oldestPending->notEmpty and
          t.allReservations->select(pending)->forall(  r
|
            t.oldestPending.madeOn.notAfter(r) ) )
-- In the context of a title t
--   if there are no pending reservations for t (check
by selecting
--   the pending ones and seeing if the resulting set
is empty) then
--       there is no oldest pending reservation
-- and
--   if there are some pending reservations for t then
--       there must be an oldest pending and
--         for all reservations, r, that are pending
for t,
--            the timestamp showing when the oldest
pending
--            reservation was made is NOT after the
timestamp of r
```

More loosely, the invariant says that, if there are some pending reservations, there must be an oldest one, and this oldest one has the property that there is not an even older one. The specification intentionally doesn't say which is the oldest pending reservation out of two with equal time stamps.
The example was chosen to illustrate how the definitions of the Boolean operators avoid a problem with expressions that have no defined meaning.
In the very last line of the formal version of the invariant, what does the subexpression "t.oldestPending" mean when there are no pending reservations? The

126     S. Cook et al.

intuition behind the formal specification is that, when there are no pending reservations, there is no "oldest pending" reservation, so the question is irrelevant. To support this intuition, the "implies" operator is defined so that "a implies b" is true whenever "a" is false - we do not need to examine "b".

However, the formal logic underpinning OCL does assign a "virtual value" to "b". This value is neither true nor false, but represents "undefined". As a consequence the Boolean type does have three values, "true", "false", and "undefined" and the Boolean operations need to be defined on "undefined" as well. Fortunately, we can use this special value to describe the desired behaviour of our operations. We e.g. can define that "false implies undefined" is true (the formal logic is known as Kleene logic). Please note, that the value "undefined" was an invention of semantics modellers (formalists) to get their understanding of the semantics of a logic like OCL right. The technique of introducing virtual values was very successful, and is nowadays often used to model certain situations, like "nil" models the absence of an object to navigate to.

As the example concerning reservations illustrates, modellers usually do not need to consider "undefined" values explicitly. They model them out by carefully exploiting definitions such as "false implies b is true, for any expression b that has type Boolean". As a result, most of the time, modellers can choose freely to work with any one of these three informal pictures of what Boolean expressions mean when they have subexpressions that cannot be given a value.

One interpretation of this situation is operational: operators such as "and" and "implies" are evaluated by a process that stops as soon as the answer is knowable (for example, when either of b1 or b2 in "b1 and b2" is found to be false).

Boolean expressions can have the values false, true or undefined, and the rules for evaluating the operators take care of undefined values by including such rules as "false and undefined = false"

In another yet consistent interpretation all Boolean expressions are either true or false, but, because the modeller has (deliberately) under-specified an expression, it might not be possible to determine which. However, rules such as "false and b = false" mean that, even if "b" is underspecified, so that its value is not known, the value of "false and b" is still known (its value is false).

When a modeller does need full, formal definitions of the Boolean operators, for example, to construct formal proofs, such definitions are available.

### 3.2     Precise Definition of the Boolean Operators on the Undefined Value

Based on the introduction to the Boolean operators in OCL, we in this section present the plain, precise facts. We describe the effect of the operators

=, not, and, or, xor, implies, if-expression

through truth tables and present some of the laws that still hold. The presented logic is called 3-valued logic in [Kleene52], and therefore often also called Kleene Logic.

### 3.2.1   The = Operator

First of all, it should be clarified, what equality on undefined values means. There are basically two version of equality. The so called "strong equality" and the "weak equality". Strong equality is capable of comparing undefined values. In particular the last line of the table allows us to check in the specification whether a value is undefined.

| b1 | b2 | b1 = b2 |
|---|---|---|
| false | false | true |
| false | true | false |
| true | false | false |
| true | true | true |
| undefined | true | false |
| undefined | false | false |
| true | undefined | false |
| false | undefined | false |
| undefined | undefined | true |

Unfortunately the strong equality cannot be implemented. In an implementation instead a weak or strict equality must be used. This is why specification languages normally in addition introduce a weak equality (e.g. named as "==") with the following properties:

| b1 | b2 | b1 == b2 |
|---|---|---|
| false | false | true |
| false | true | false |
| true | false | false |
| true | true | true |
| undefined | true | undefined |
| undefined | false | undefined |
| true | undefined | undefined |
| false | undefined | undefined |
| undefined | undefined | undefined |

Fortunately both versions fulfil the equality law:

| b1 = b2 | is equal to | b2 = b1 |

| b1 == b2 | is equal to | b2 == b1 |

Please note that the strong equality is defined for the undefined values in all other types as well. Eg. "1/0 = sqr(-1)" yields "true" as both sides are equal to the undefined value of type Real.

Currently OCL does not have both versions of equality. The strong equality is important to deal with undefined values (e.g. to describe that a certain value is not undefined). In contrast, weak equality can be defined like a normal function. So we choose "=" to be the strong equality.

### 3.2.2  The NOT Operator

| b | not b |
|---|---|
| true | false |
| false | true |
| undefined | undefined |

It remains valid that:

  not not b=b

### 3.2.3  The AND Operator

| b1 | b2 | b1 and b2 |
|---|---|---|
| false | false | false |
| false | true | false |
| false | undefined | false |
| true | false | false |
| true | true | true |
| true | undefined | undefined |
| undefined | false | false |
| undefined | true | undefined |
| undefined | undefined | undefined |

Some basic laws for the and operator:

| commutative: | ( b1 and b2 ) | = | ( b2 and b1 ) |
| associative: | ( (b1 and b2) and b3) | = | ( b1 and (b2 and b3) ) |
| false is dominant: | ( b and false ) | = false | =(false and b) |
| true is neutral: | ( b and true ) | = b = | (true and b) |
| idempotence: | ( b and b) | = | b |

Note that although the and operator sometimes returns undefined values on undefined arguments, this operator can be implemented. However, both arguments need to be evaluated in parallel until the first return the value false.

### 3.2.4   The OR Operator

| b1 | b2 | b1 or b2 |
|---|---|---|
| false | false | false |
| false | true | true |
| false | undefined | undefined |
| true | false | true |
| true | true | true |
| true | undefined | true |
| undefined | false | undefined |
| undefined | true | true |
| undefined | undefined | undefined |

Some basic laws for the or operator:

| commutative: | ( b1 or b2 ) | = | ( b2 or b1 ) |
|---|---|---|---|
| associative: | ( (b1 or b2) or b3 ) | = | ( b1 or (b2 or b3)) |
| true is dominant: | ( b or true ) | = true = | ( true or b ) |
| false is neutral: | ( b or false ) | = b = | ( false or b ) |
| idempotence: | ( b or b) | = | b |

### 3.2.5   The XOR Operator

The xor operator ("exclusive or") holds if exactly one of its arguments holds. It is therefore similar to the or operator, but excludes the case that both arguments are true. Therefore the xor operator relies more on the values of its arguments:

The xor operator is defined in terms of the and, or and not operators.

(b1 xor b2)          =          ( (b1 or b2) and not (b1 and b2) )

From this definition, we derive the following truth table:

| b1 | b2 | b1 xor b2 |
|---|---|---|
| false | false | false |
| false | true | true |
| false | undefined | undefined |
| true | false | true |
| true | true | false |
| true | undefined | undefined |
| undefined | false | undefined |
| undefined | true | undefined |
| undefined | undefined | undefined |

Some basic laws for the xor operator:

| | | | |
|---|---|---|---|
| commutative: | ( b1 xor b2 ) | = | ( b2 xor b1 ) |
| associative: | ( (b1 xor b2) xor b3 ) | = | ( b1 xor (b2 xor b3) ) |

### 3.2.6 The IMPLIES Operator

The implies operator is defined in terms of the and, or and not operators.

    ( b1 implies b2 )    =    ( (not b1) or b2 )

| b1 | b2 | b1 implies b2 |
|---|---|---|
| false | false | true |
| false | true | true |
| false | undefined | true |
| true | false | false |
| true | true | true |
| true | undefined | undefined |
| undefined | false | undefined |
| undefined | true | true |
| undefined | undefined | undefined |

Some basic laws for the implies operator:

  ( b1 implies (b2 implies b3) )  =    ( (b1 and b2) implies b3 )

The implies operation has a lot of uses. E.g. the actual postcondition of a method is given by "pre implies post", provided that "pre" and "post" are the explicitly given conditions. Unfortunately the standard rule ( b implies b ) = true does not hold for b = undefined.

Please note that there exists no symmetric equivalent "<=>" or "bi-implies" in OCL. Instead the strong equality "=" must be used to describe that two Boolean expressions are equal. This is a bit awkward because "=" has a higher precedence than "<=>" would have. Therefore in the above laws we needed to set a number of extra brackets.

### 3.2.7 The IF-Expression

The if expression can be charactericed by the following three equations:

| | | |
|---|---|---|
| if true then x else y endif | = | x |
| if false then x else y endif | = | y |
| if undefined then x else y endif | = | undefined |

resembling exactly the standard if-then-else-semantics. A law for the if then else:

| | | |
|---|---|---|
| if b1 then x else y endif | = | if (not b) then y else x endif |

### 3.3 Clarify the Meaning of Recursive Definitions

Recursion always occurs if an element refers to its own definition while being defined. Functions and methods can be recursive. Famous examples are the factorial function:

**fac**(n) = if (n==1) then return(1) else return(n* **fac**(n-1));

Or list traversals as e.g. used in container classes:

list.**length**() = if (list.elem==NIL) then return(0) else return 1 + **length**(list.next);

But structures may also be recursive, for example by including an attribute in a class that refers to the class itself (used in list implementations) or to a superclass (used in part-whole-structures). Also OCL constraints may use recursive definitions, like the following:

```
context Person invariant:
ancestors = parents->union(parents.ancestors)
```

Which is defined in the context of the following class:

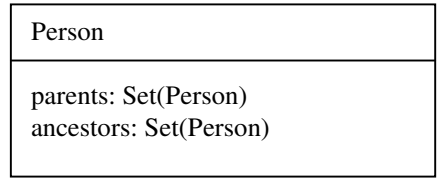

In the above constraint ancestors is meant to be the transitive closure of the relation defined by parents. Reading the constraint as a functional program, then we actually have defined what we wanted to.

When dealing with recursive specifications instead of programs, then there are typically more than one solutions. This can be easily seen regarding the OCL constraint

a = a*a - 4*a + 6

This is not a valid mathematical definition for a, but a binomial equation with two solutions (a=2 or a=3). Both are valid and none is preferable. The same situation arises with the ancestors. Let's look at the following object structure, where Joe is Mary's dad and Pete's grandpa:

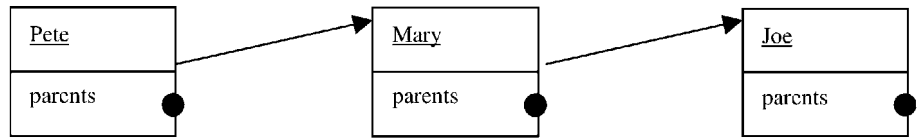

The desired solution is covered by the following object structure (disregarding the parents attribute):

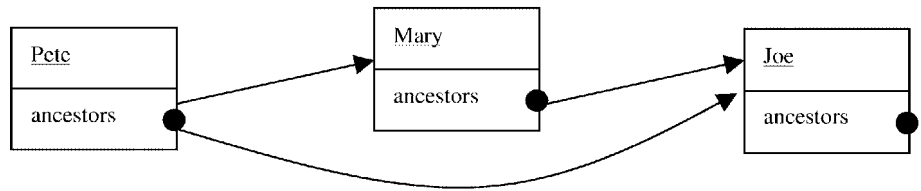

But there are more solutions to the ancestor constraint as shown in the following object structure:

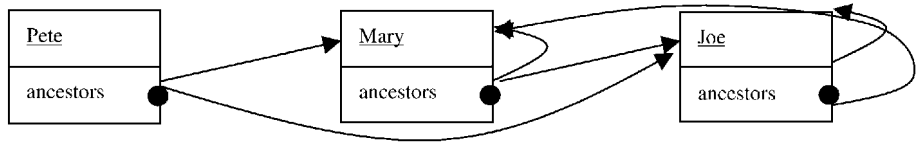

It demands that Mary and Joe are both ancestors of each other and themselves. Through checking the constraint on the ancestors attribute, we find that both shown object diagrams are valid structures. And there are a lot more.

Fortunately, there is a unique characterisation of all possible object structures that furthermore distinguishes one solution from all others. The first object structure shown, is contained in all others, in the sense that each solution must contain at least the link from Pete to Mary, from Pete to Joe and from Mary to Joe. This solution is "minimal" and therefore corresponds to the "minimal fixpoint" and is identical to what we get, if "executing" the ancestor definition as if it would be a program.

Unfortunately, such a minimal solution does not always exist, sometimes there are several ones, sometimes none at all. Even worse, it may be difficult to find out, whether a recursive definition has a unique solution. For example the equation about "a" from above does have two solutions, but none of it can be preferred in the domain-theoretic sense. This is also reflected, if you try to "execute" the equation, by iterative application. The sequence

$a_{n+1} = a_n * a_n - 4 * a_n + 6$

starting with, say $a_n = 0$, leads to a limit of $+\infty$ and therefore no solution at all.

It may, furthermore, happen that the modeller does not want to get the minimal solution. Consider a specification, where Person stands for computer nodes, "parents" is replaced by "TCP-connection" and "ancestors" is the transitive closure of all "connections", describing to whom a connection can be made. Then the constraint tells us, that connection includes all TCP-connection, but other kinds of connections may exist, that want to add connections later.

By the way, these considerations have been explored fully with the fixpoint theory. Fixpoint techniques, like the one provided by Kleene [Kleene52] and Tarski [Tarsky55] give an unambiguously defined semantics for recursive definitions. If fixpoint theory is familiar, then you know about the results discussed above. If not, then just accept the results presented here, as you probably don't want to bother with the technical issues here. From first order logic languages, we know how to deal with that situation. There are two approaches possible, both have their merits and problems.

One approach is to use "loose semantics": Only the constraints are valid that are explicitly given or can be derived from given constraints. No implicit "minimal solution" semantics is used, when recursive equations occur. If a minimal solution is desired, a special keyword or stereotype is to be added to the constraint. The keyword could be "executable" or "operational" to give a hint to the operational semantics of recursive definition.

Another approach is to use the minimal solution as default and provide a keyword, like "loose" to indicate that any solution is possible.

Although the first approach is somewhat cleaner, as it does not add implicit constraints, the second one is probably more practical. A solution can be to include both kinds of keywords to allow to explicitly mark which kind of semantics is to be used.

**Some technical remarks on recursion**. Recursion can be more complicated through involvement of several elements, like in

```
context Person invariant:
grandancestors = parents.ancestors;
ancestors = parents->union(grandancestors)
```

where the recursion is mutually dependent between ancestors and grandancestors, but not direct. But the standard keywords can be extended to such cases.

The nature of first order logic languages does not allow to uniquely characterise a minimal solution. This is only possible by adding second order principles, such as induction or term generation of the specified elements. Usually introducing special keywords (like the ones proposed above) provides such principles. Although we recommend not to use the technique (see Sec. 2.3) we can use the built in OCL Integers and their induction principle built in and therefore provides the necessary techniques. It is a bit awkward, but the minimal solution can be specified through using natural numbers and explicitly mimicking induction over natural number n:

| Person |
| parents: Set(Person) |
| ancestors: Set(Person) |
| ancestors-up-to(n:Nat): Set(Person) |

```
context Person invariant:
ancestors-up-to(n) =
    if (n==1) then parents else
         parents->union(parents.ancestors-up-to(n-1))
Integer->forall(n | ancestors-up-to(n) = ancestors-up-
to(n+1) and n>=1
     implies ancestors = ancestors-up-to(n) )
```

The use of an appropriate keyword is absolutely preferred.

From our experience using OCL for specification, we found that recursive situations frequently occur. Particular sources are recursive data structures (lists, containers, part-wholes, directory structures, hierarchical, and graph-containing structures). Usually recursion of OCL constraints is accompanied with the existence of an association circle in the class diagrams. When one or several associations form a circle, building paths through them can lead to recursive structures.

### 3.4    Use Path Concept Only for Packages

The UML and OCL specification uses a double colon "::" to denote pathnames. This is used e.g. for denoting Classes in other packages as with "Packagename:: Classname". The same construct has been used in OCL to denote the use of a redefined feature. Because this uses the same notation for something else than a pathname, it was considered to make the specification less consistent.
In UML 1.3 the use of "::" to access redefined features has been removed. Instead, one has to use the "oclAsType" operation to cast an object to its supertype and then access the redefined feature.

### 3.5    Other Subjects That Should Be Clarified

During our meeting we discussed other subjects in the OCL standard that needed clarification. However, up to September 1999 we did not have the opportunity to discuss them further and include a clarification in this manifesto. The subjects were:

- Explain navigation paths like "a.b.c" without flattening
- Precisely define the scope of invariants
- Type expressions in OCL
- Extensions to OCL to make it more useful

### 3.6    New Operator "oclIsNew" for Postconditions

#### 3.6.1    Rationale:

One of the most common ways to denote the existence of new objects was the use of "allInstances". Because this operation is not well-defined (see section 2.3) there should be another way to denote that an instance has been created. The operation "oclIsNew" has been added to the OclAny type for this purpose.

### 3.6.2   Syntax:

The syntax of "oclIsNew" in UML 1.3 is as follows:
```
object.oclIsNew : Boolean
```

### 3.6.3   Semantics:

In UML 1.3 the operation "oclIsNew" can only be used in a post condition. It evaluates to true if the *object* is created during performing the operation. I.e. it didn't exist at precondition time.

## 3.7   New Operator "isUnique"

### 3.7.1   Rationale:

In practice one often wants to ensure that the value of an attribute of an object is unique within a collection of objects. The isUnique operation gives the modeller the means to state this fact. This operator is now part of the OCL 1.3.

### 3.7.2   Syntax:

```
collection->isUnique(expr : OclExpression) : Boolean
```

### 3.7.3   Semantics:

The isUnique operation returns true if "expr" evaluates to a different value for each element in the collection, otherwise the result is false.
More formally:
```
collection->isUnique(expr : OclExpression)
```
is identical to
```
collection->forAll(e1, e2| if e1 <>e2 then e1.expr <>
e2.expr)
```

### 3.7.4   Example Usage:

```
context LoyaltyProgram invariant:
serviceLevel->isUnique(name)
```

### 3.8 Add a "let" Statement to Define Local Variables and Functions

#### 3.8.1 Rationale:

For larger constraints it is cumbersome and error prone to repeat identical sub expressions. The possibility to define a local variable within a constraint solves this problem.
Sometimes a sub-expression is used more than once in a constraint. The *let* expression allows one to define a variable, which can be used in the constraint.

#### 3.8.2 Syntax:

The syntax of a Let expression is exemplified as follows:

```
context Person invariant:
let income : Integer = self.job.salary->sum in
if isUnemployed then
    income < 100
else
    income >= 100
endif
```

#### 3.8.3 Semantics:

The variable defined in a let statement can be used within the OCL expression. The value of the expression is substituted for the variable. Because semantically nothing can change during evaluation of a constraint, it is not necessary to define whether the expression at the left hand side is evaluated once at the start, or at each occurrence in the OCL expression.

### 3.9 Introduce a Read-Only Modifier for Attributes[1]

#### 3.9.1 Rationale

In current UML, attributes are either public, protected, or private. These modifiers constrain visibility of attributes to foreign classes, subclasses and to other methods of the same class. Whenever an attribute is visible, it can be both read and changed. Sometimes, it is useful to constrain the visibility to be readable, but not changeable. For example, current time, a counter, or a reference to certain objects could be read-only to the environment, but writable to the own methods. Currently such values must be private and used through (rather expensive) method access. We therefore introduce the concept of a read-only-modifier in combination with private and protected.

---

[1] Note, that this is actually a proposal for extending UML.

138    S. Cook et al.

### 3.9.2    Syntax

public read  attributename
public read   protected write attributename
protected read attributename

### 3.9.3    Semantics

In addition to the existing three modifiers, we introduce the above given modifier combinations with the following meaning:

| | |
|---|---|
| public read | read access is public, write is private (!) |
| public read   protected write | read access is public, write is protected |
| protected read | read access is protected, write is private |

Write access is always at least as constrained as read access. The three variants, where the write access is more constrained than the read access, are covered above. Shortcuts, like "pubread", "protread" and "protwrite" are possible, but not very elegant.

### 3.9.4    Example Usage:

| Person |
|---|
| public read parents: Set(Person)<br>protected read ancestors: Set(Person) |

## 3.10    Constant Declarations for Object Properties

### 3.10.1    Rationale

Some properties of objects never change. Marking them to show this allows for some additional checking, e.g. such properties can never be mentioned as changing in a postcondition, and also allows for more reasoning to be done over the object model.

### 3.10.2 Syntax

```
context object invariant:
constant <attribute>
constant <query()>
constant <rolename>
```

### 3.10.3 Semantics

Where an attribute or query has been declared as constant, no postcondition can be specified which implies any change to the value of that attribute or query, unless the postcondition also states that the object to which that attribute or query is applied is new.

Where an association end (role name) has been declared as constant, no postcondition can be specified which implies any change to the collection of objects denoted by that role name, unless the postcondition also states that the object to which that role name is applied is new.

Note that this declaration relates to the attribute 'changeability' of StructuralFeature (superclass of Attribute), where one of the values is *frozen*. The class AssociationEnd (rolename) has the same attribute in the UML metamodel, but Operation (query) is lacking this attribute.

### 3.10.4 Example Usage

```
context Customer invariant:
constant dateOfBirth
```

## 3.11 Enhance the Abilities to Define Context of OCL Specifications

Expressions written in Object Constraint Language (OCL) within a UML model assume a context, depending upon where they are written. In UML 1.3 the exact nature of this context is not fully defined. Furthermore there is no mechanism for defining the context for OCL expressions in extensions to UML. In a separate paper [Cook99a] the context of OCL expressions is defined and a precise and flexible mechanisms for how to specify this context is proposed.

## 3.12 Other Extensions That Were Discussed

During our meeting other extensions and improvements were discussed:
- Useful syntactic abbreviations for existing constructs
- Add "effect" definitions (from Catalysis) for postconditions

## 4 Application of OCL

The following issues demonstrate different uses and extensions of OCL towards increasing its expressiveness. These are just suggestions whose value needs to be carefully examined and the proposed concepts improved accordingly.

### 4.1   Using OCL to Define Events, Operations, and Actions

#### 4.1.1   Background and Rationale

UML/OCL allows us to specify operations using preconditions and postconditions. Catalysis introduces the idea of joint actions. A key difference between operations and joint actions is that operations are *localised* on a single type, whereas joint actions are joint over two or more types. UML/OCL should explicitly embrace joint actions, and to go even further, allow *events,* which are not localised on any types at all.
Note that there's nothing original here. The ideas presented here come from Syntropy [Cook94] and Catalysis [D'Souza99].

#### 4.1.2   Operations

Users of object technology are familiar with the idea of what Smalltalk and Java call a method, C++ calls a member function, Eiffel calls a routine, and UML calls an operation. Those familiar with Syntropy, Fusion, OCL, Eiffel, etc. will know about using preconditions and postconditions to specify the behaviour of an operation.
The general form of an operation specification in OCL is this:

```
context  Typename::operationName(parameter1  :  Type1,
... ): ReturnType
  pre :  parameter1 ...
  post:  result = ...
```

Within the assertions labelled *pre:* and *post:* the term *self* refers to an object of type *Typename*. Note that if the return type is void, it can be omitted.
Example
In a model of a library, there could be an operation *borrow* on type *Library*, specified along these lines (assume that every library object has a clock object):

```
operations Libary::borrow( m : Member, c : Copy, d :
Date )
  -- Member m borrows copy c to be returned on date d
```

```
    pre:
       -- The date is in the future
       d > self.clock.today
          ...
    post:
         -- There's a new loan object recording that m has
borrowed c
         Loans.allInstances -> exists( n | n.isNew and ... )
```

The operation is localised on the library. It is natural to think of the operation *borrow* being called on a library object, or of a library object "receiving" a call to its *borrow* operation. The term *self* in the precondition refers to the library that "receives" the call.

### 4.1.3    Joint Actions

In Catalysis, we can have joint actions. They are joint in the sense that they are localised on two or more types.
Example
The borrow action can be seen as a joint action between a member and a library system.

```
            action (m : Member, lb : Libary)::borrow( c : Copy, d :
            Date )
                 -- A joint action in which a member and a
            library system
                 -- collaborate in the borrowing of a copy c by
            the member
                 -- from the library, to be returned on date d
            pre:
                 -- The date is in the future
                 d > lb.clock.today
                 ...
            post:
                 -- There's a new loan object recording that m has
            borrowed c
                 Loans.allInstances -> exists( n | n.isNew and ... )
```

Note that the library is now identified by name. The term *self* is no longer unambiguous.

### 4.1.4     Events

An operation is localised on a single type. A joint action is less localised. We don't need to think of borrowing as an operation on a library. We can think of it as a piece of behaviour that a member and a library collaborate in performing.

We can go further, and have no localisation at all. We call a fully de-localised operation an *event*, following Syntropy.

Example

The borrowing of a copy by a member from a library, with a certain return date, can be seen as an event that involves a number of objects. The event has a before state and an after state, but is not done *to* any particular object. Rather, it can affect any of the objects identified in the event's signature, and any object that can be reached from them.

```
event borrow( lb : Library, m : Member, c : Copy, d :
Date )
        -- An event in which member m borrows copy c
from library l,
        -- to be returned on date d
pre:
    -- The date is in the future
    d > lb.clock.today
    ...
post:
    ...
```

There is no receiving object, and nothing for the term *self* to refer to.

More strictly, what is specified above is an event type. Any event occurrence will identify instances of types Library, Member, etc.

## 4.2     Adding Dynamics to OCL

### 4.2.1     Rationale

Currently OCL expressions state only static requirements. Class invariants are static by nature and even guard conditions, pre- and postconditions express static information. They only constrain changes in the object's state. Although this is very useful, one would often like to express more dynamic constraints.

What can not currently be expressed using OCL is, for instance, what should happen at the exact moment that a condition becomes true. Another example is, the case where one would like to express that although the state of the object is not changed, events must have occurred due to the invocation of the current operation.

### 4.2.2    Proposed Features

Special features need to be introduced in OCL to be able to specify these dynamic constraints. We propose to add two stereotypes for constraints to UML and their syntax to OCL that will support dynamic constraints:
the **action** stereotype, which indicates a constraint on the triggering of messages, and
the **called** stereotype, which indicates a constraint that ensures certain messages are send.
To explain these new features the example UML model shown in Figure 2 will be used.

### 4.2.3    The Action Stereotype for Constraint

An action constraint states that when a Boolean expression becomes true, a list of messages must be send. The constraint can be used to guarantee the sending of the messages, but not the completion of the invoked operations. The messages in the messagelist can only be sent to the contextual object or objects navigable from the contextual object.

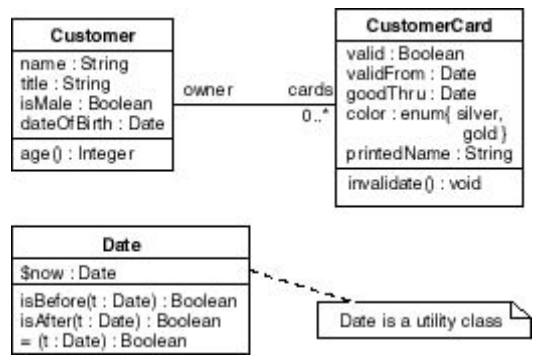

**Fig. 2.** Another example class diagram

#### 4.2.3.1    Syntax:

```
context class action:
    on booleanExpression do messagelist
```

Where booleanExpression is an OCL expression of the boolean type and messagelist is defined as featureCall ("," featureCall)* (featureCall as defined in the formal grammar of OCL).

#### 4.2.3.2    Semantics:

To define the semantics of an action constraint we propose that an action constraint has the same semantics as an automatic, conditional transition in a statechart. Figure 3 shows a statechart with the same meaning as the example below.

#### 4.2.3.3    Example Usage:

```
context CustomerCard action:
on self.goodThru.isAfter(Date.now) do
self.invalidate()
```

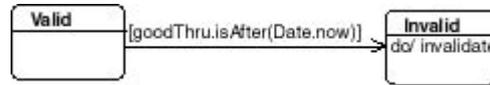

**Fig. 3.** Statechart that resembles the action constraint

### 4.2.4    The Called Stereotype for Constraint

A called constraint can only be used in the context of an operation. A called constraint states that every time the contextual operation is executed a message or a list of messages is being sent. Again, there is no guarantee that the invoked operations are completed. The messagelist may be included in ordinary OCL-expressions, allowing e.g. if-then-else-clauses. The else branch may be empty, in which case the keyword 'else' may be omitted.

#### 4.2.4.1    Syntax:

```
context class.operation(paramlist): resultType
called: messagelist
```

#### 4.2.4.2    Semantics:

The semantics of the called constraint can be defined in terms of sequence diagrams. Stating that a message Y has been send during the execution of an operation X is equal to requiring that in every sequence diagram where operation X is called, this call is followed by message Y. In the case of a conditional message, the call X is followed by a guard and message Y.

#### 4.2.4.3    Example Usage:

```
context CustomerCard::invalidate(): void
pre:    -- none
post:   valid = false
called: if valid@pre = true then
            if customer.special then
customer.sendPoliteInvalidLetter()
            else customer.sendInvalidLetter()
            endif
        endif
```

#### 4.2.5   Consequences of Introducing Dynamics

What is described above is a proposal of which the consequences need to be examined. Some of these are:
As in pre and postconditions we do need some formalisation of the concept 'time'. Perhaps we only need to specify 'moment', 'before' and 'after'.
There surely will be a relation to the proposed UML action language specification, which needs to be investigated further.

### 4.3   Mapping STD to Object States Using OCL

#### 4.3.1   Rationale

A state machine for a Classifier defines the states in which an instance of the classifier can be. It is common to use the state of an object in preconditions, postconditions and invariants. Therefore, it is useful to enhance OCL to be able to refer to those states.

#### 4.3.2   Syntax

The standard operation "oclInState" has been added to OclAny.
```
          object.oclInState(state : OclState) : Boolean
```

#### 4.3.3   Semantics

Results in true if *object* is in the state *state*, otherwise results in false. The argument is a name of a state in the state machine corresponding with the class of *object*.
The type OclState is used as a parameter got the operation "oclInState". There are no properties defined on OclState. One can only specify an OclState by using the name of the state, as it appears in a state machine. These names can be fully qualified by the nested states and the state machine that contain them.

### 4.4   Using Prefaces to Customise UML

The UML is extensible, and so can be regarded as a family of languages. Implicitly or explicitly, any particular UML model should be accompanied by a definition of the particular UML family member for the model. The definition should cover syntactic and semantics issues. A preface is a mechanism for associating models with such definitions.
This topic was further elaborated on in an article to be published in the proceedings of the Tools Pacific 1999 conference. Readers are referred to [Cook99b].

### 4.5     Explicit Import of UML and OCL Metamodels Used to Customise UML

Often we generate OCL expressions as strings. For a thorough treatment, this is not a satisfying approach. Although OCL is not a graphic language, it has an abstract syntax and can therefore be included in the meta-model approach in quite the same way as other UML diagrams are. Figure 4 contains a subset of an OCL preface, a similar, but much more detailed approach has been used in [Richters99].

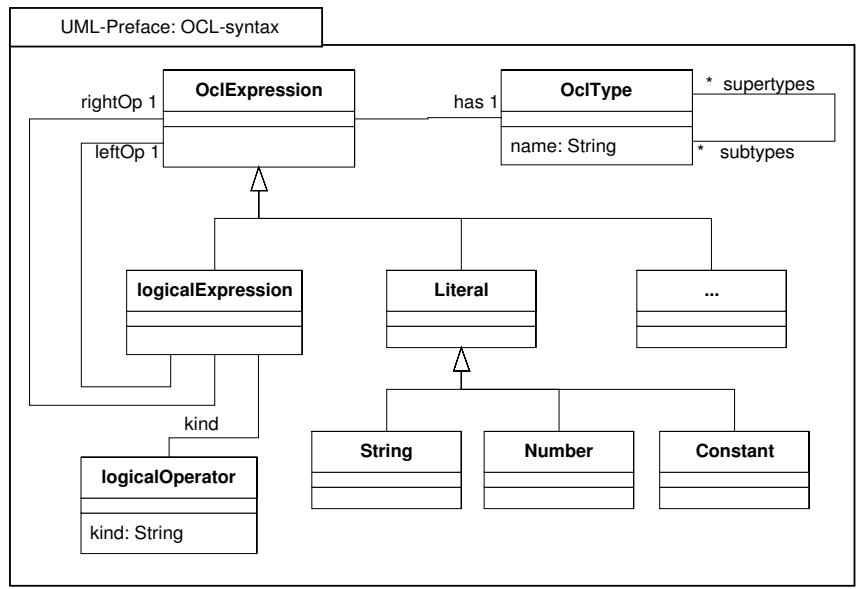

**Fig. 4.** UML-Preface package for OCL (subset)

Using the OCL preface allows manipulating OCL expressions in the same way as ordinary class structures. OCL on the meta-level can be used to constrain OCL expressions on the modelling level, without the usual conflicts, which arise when two levels are mixed. In particular, OCL has a strong type system, both on the meta-level and the level below. However, through manipulating OCL expressions on the meta-level, we can easily construct OCL expressions on the modelling level that are not well formed. It is not as easy as it was when using strings, but still e.g. variables may be used that do not exist or are of a wrong type.

To deal with that issue, the class "OclExpression" can offer an appropriate query, e.g. "correctExpr()" that discovers well-formedness errors of OCL expressions on the modelling level. The type "OclExpression" as part of OCL's own type system is not necessary anymore.

### 4.6 Other Subjects Discussed

- Expression placeholders
- Define a basic inference calculus
- How to deal with exceptions in OCL

**Acknowledgements.** We would like to thank Mark Richters for his comments on an earlier version of the paper.

## References

| | |
|---|---|
| [Bauer85] | F. L Bauer, R. Berghammer, M. Broy, W. Dosch, F. Geiselbrechtinger, R. Gnatz, E. Hangel, W. Hesse, B. Krieg-Brückner, A. Laut, T. Matzner, B. Möller, F. Nickl, H. Partsch, P. Pepper, K. Samelson, M. Wirsing, H. Wössner. The Munich Project CIP, Vol 1: The Wide Spectrum Language CIP-L. Springer Verlag, LNCS 183, 1985 |
| [Booch98] | G. Booch, J. Rumbaugh, and I. Jacobson. The Unified Modelling Language User Guide. Addison Wesley Longman, Reading, Massachusetts, 1998 |
| [Broy93-1] | M. Broy, C. Facchi, R. Grosu, R. Hettler, H. Hussmann, D. Nazareth, F. Regens-burger, O. Slotosch, K. Stoelen. The Requirement Design Specification Language. An Informal Introduction, Part 1. Spectrum. TUM-I9312. TU Munich. 1993 |
| [Broy93-2] | M. Broy, C. Facchi, R. Grosu, R. Hettler, H. Hussmann, D. Nazareth, F. Regens-burger, O. Slotosch, K. Stoelen. The Requirement Design Specification Language. An Informal Introduction, Part 2. Spectrum, TUM-I9312, TU Munich, 1993 |
| [Cook94] | S. Cook and J. Daniels, Designing Object Systems, object-oriented modelling with Syntropy, Prentice Hall, 1994 |
| [Cook99a] | S. Cook, A. Kleppe, R. Mitchell, J. Warmer, A. Wills, Defining the Context of OCL Expressions. In: <<UML>>'99 The Unified Modelling Language. Eds.: R. France, B. Rumpe. Springer Verlag, LNCS 1723, 1999. |
| [Cook99b] | Steve Cook , Anneke Kleppe, Richard Mitchell, Bernhard Rumpe, Jos Warmer, Alan Wills, Defining UML Family Members with Prefaces, Proceedings of Tools Pacific '99, 1999 |
| [Conrad92] | S. Conrad, M. Gogolla, R. Herzig. TROLL light: A Core Language for Specifying Objects. Technical Report 92-06. TU Braunschweig. 1992 |
| [Date87] | J. C. Date. A Guide to SQL Standard. Addison Wesley. 1987 |
| [D'Souza98] | D'Souza D. and Wills A. Objects, Components and Frameworks with UML: The Catalysis Approach. Addison Wesley, 1998 |

[Ehrig85]         H. Ehrig, B. Mahr, Fundamentals of Algebraic Specification 1, Springer Verlag, 1985

[Ehrig90]         H. Ehrig, B. Mahr, Fundamentals of Algebraic Specification 2, Module Specifications Constraints, Springer Verlag, 1990

[Goguen92]        J. Goguen, T. Winkler, J. Meseguer, K. Futatsugi, J.-P. Jouannaud. Introducing OBJ. Technical Report CSL-92-03. Computer Science Laboratory, SRI. 1992

[Gordon93]        M. Gordon, T. Melham. Introduction to HOL: A Theorem Proving Environment for Higher Order Logic. Cambridge University Press. 1993

[Gordon79]        M. Gordon, R. Milner, C. Wadsworth. Edinburgh LCF: A Mechanised Logic of Computation. Springer Verlag, LNCS 78. 1979

[Hartmann94]      T. Hartmann, G. Saake, R. Jungclaus, P. Hartel, J. Kusch. Revised Version of the Modelling Language TROLL. Technical Report 94-03. TU Braunschweig. 1994

[Hudak92]         P. Hudak, S. P. Jones, P. Wadler. Report on the Programming Language Haskell. A Non-strict Purely Functional Language. Sigplan Notices. Vol. 27. ACM Press. 1992

[Jones90]         C. B. Jones. Systematic Software Development Using VDM. Prentice Hall. 2$^{nd}$ Edition. 1990

[Jones93]         M. P. Jones. An Introduction to Gofer. 1993

[Jungclaus91]     R. Jungclaus, G. Saake, T. Hartmann, C. Sernadas. Object-oriented Specification of Information Systems: The TROLL Language. TU Braunschweig. Technical Report 91-04. 1991

[Lalonde90]       W. Lalonde and J. Pugh, Inside Smalltalk, vol 1, Prentice Hall, 1990

[Kleene52]        S. Kleene. Introduction to Metamathematics. Van Nostrand. 1952

[OCL1.3]          OMG Unified Modeling Language Specification Version 1.3 beta R7, June 1999

[OCL1.4]          OMG Unified Modeling Language Specification Version 1.4, to appear 1999/2000.

[Paulson91]       L. Paulson. ML for the Working Programmer. Cambridge University Press. 1991

[Paulson94]       L. Paulson. Isabelle: A Generic Theorem Prover. Springer Verlag, LNCS 929. 1994

[Regensburger94]  F. Regensburger. Phd Thesis. HOLCF: Eine konservative Erweiterung von HOL um LCF. TU Munich. 1994

[Richters99]      M. Richters, M. Gogolla: A Metamodel for OCL. In: <<UML>>'99 The Unified Modelling Language. Eds.: R. France, B. Rumpe. Springer Verlag, LNCS 1723, 1999.

[Rumpe98]         B. Rumpe: A Note on Semantics (with an Emphasis on UML). In: Second ECOOP Workshop on Precise Behavioral Semantics. Technical Report TUM-I9813. Technische Universität München. Juni, 1998

[Spivey88]        J. Spivey. Understanding Z. Cambridge University Press. 1988

[Spivey89]        J. Michael Spivey. An Introduction to Z and Formal Specifications. IEE/BCS Software Engineering Journal, vol. 4, no. 1, pp 40-50. 1989

[Tarski55]      A. Tarski. A lattice-theoretical fixpoint theorem and its application. Pacific Journal of Mathematics, vol. 5, pp. 285-309. 1955

[Warmer99a]     Warmer J. and Kleppe A. The object constraint language. Precise modelling with UML. Addison Wesley Longman, 1999

[Warmer99b]     Warmer J. and Kleppe A. OCL: the constraint language of the UML. In the Journal of Object-Oriented Programming, May 1999.